\documentclass[prl,aps,amsmath,showpacs,amsfonts,amssymb,twocolumn]{revtex4}
\usepackage{epsfig}
\usepackage{bm}
\begin{document}
\title{Driven Pair Contact Process with Diffusion}
\author{Su-Chan Park}
\affiliation{School of Physics, Korea Institute for Advanced Study, Seoul 130-722, Korea}
\author{Hyunggyu Park}
\affiliation{School of Physics, Korea Institute for Advanced Study, Seoul 130-722, Korea}
\date{\today}
\begin{abstract}
The pair contact process with diffusion (PCPD)  has been recently
investigated extensively, but its critical behavior is not yet clearly 
established. By introducing biased diffusion, we show that the external driving
is relevant and the driven PCPD exhibits a mean-field-type critical behavior
even in one dimension.
In systems which can be described by a single-species bosonic field theory,
the Galilean invariance guarantees that the driving is irrelevant. 
The well-established
directed percolation (DP) and parity conserving (PC) classes are such examples.
This leads us to conclude that the PCPD 
universality class should be distinct from 
the DP or PC class. Moreover, 
it implies that the PCPD is generically a multi-species model 
and a field theory of two species is suitable for proper description. 
\end{abstract}
\maketitle

The steady state of an equilibrium system is 
characterized by its Hamiltonian and Gibbs measure. 
There is no systematic generalization to the stationary state of 
nonequilibrium systems so far. Since nonequilibrium systems 
encompass all kinds of many body systems without a constraint of 
detailed balance, it may be hopeless to find a universal formalism 
applied to general nonequilibrium systems. At this point, 
model studies or case-by-case studies are rather useful  
to accumulate our knowledge on nonequilibrium systems. 

Our experience on equilibrium systems illustrates the scale-free fluctuation 
or  power law behavior at the critical point where the 
continuous phase change occurs. 
The scale-free nature is worth while to be studied not only because of its 
theoretical attraction, but also because of ubiquity in nature -- the clustering
of galaxies \cite{P80}, $1/f$ noise \cite{BTW87}, percolation structure
\cite{SA92}, to name only a few. 
This scale-free nature is also expected at criticality under 
nonequilibrium circumstances.
As a prototype of nonequilibrium critical phenomena, 
absorbing phase transitions (APTs)
have been studied extensively \cite{H00}.
APT is a transition from an active phase
to an absorbing phase in nonequilibrium steady states.
The absorbing states are defined as the configurations where
the system cannot escape by the prescribed dynamic rules.
As in equilibrium systems, this transition is possible only at 
the thermodynamic limit
because the finite systems eventually fall into the absorbing states. 

In epidemiology, for example,
the virus extinct state is an absorbing state.
Actually, the disease spreading is modeled and dubbed the contact process (CP)
by Harris \cite{H74}.
CP shows a phase transition from the virus infested state (active state),
to the quiescent state (absorbing state).
This transition is known to belong to the directed percolation (DP)
universality class.
Actually, many types of models belong to the DP class and it is conjectured
that a phase transition occurred in a system with a single absorbing state
should share the critical behavior with the DP \cite{J81,G82}. 

As in equilibrium critical phenomena, a symmetry or conservation
may play an important role in determining the universality class.
Accordingly, many nonequilibrium systems with
symmetric absorbing states or conservation laws are studied.
As expected, the additional symmetry or conservation brings forth
a series of new universality classes. Unfortunately,
the absorbing states with higher symmetry or complex conservation 
are usually unstable with respect to an infinitesimal activity 
even in one dimension. Therefore, it is difficult to find a nontrivial
scaling other than the mean-field  type, except for a few well-established
universality classes like the DP and the directed Ising or
the parity-conserving (PC) classes.

In this context, the critical behavior of the pair contact process with diffusion (PCPD) \cite{pcpdrv}
is rather surprising. Although the PCPD has no symmetry in absorbing states
and no conservation law, the PCPD seems to form a new universality class.
Actually, some authors asserted that PCPD eventually flows 
into the DP fixed point after a huge crossover time \cite{BC03}. 
However, the extensive numerical experiments \cite{KC03} indicate that
the PCPD belongs to a new universality class other than the DP or the PC.
In addition, the long-term memory present in the PCPD has been suggested 
as a source for this new universality class \cite{NP04}. 
Nevertheless, the universality issue on the PCPD is still in hot controversy 
and it is not yet clearly settled down~\cite{pcpdrv}.  There have been
some analytic attempts to analyze the PCPD through a single-species bosonic field theory,
but no satisfactory results have appeared as yet~\cite{HT97}.

In this Letter, we introduce external driving (biased diffusion) 
in various models including the PCPD and numerically observe its effect 
on the critical scaling. The external driving may serve as 
a crucial test on the universality class of the general 
absorbing-type models and also reveal important features of their critical 
scaling.  With this test, we show later that the PCPD class should 
be distinct from the DP or the PC class and the PCPD is generically 
a two-species reaction-diffusion model.

The role of driving is usually irrelevant in single-species 
reaction-diffusion systems with absorbing states (SRDA). 
The simplest examples are the pair annihilation/coagulation models
represented by $2A\rightarrow 0$ or $2A\rightarrow A$. These models can 
be solved exactly even with
biased diffusion which turns out to be irrelevant to the long time decay 
dynamics of the particle density \cite{S00}. 

In the field theoretical sense, it is easily predictable within the bosonic
formalism  introduced by Doi, Grassberger, and others \cite{bosonFormalism}. 
Since the particle density is so low in the long time limit, 
it would not be harmful to adopt the bosonic formalism where multiple 
occupations are allowed at a site. After taking a suitable
modification of the dynamic rules for bosonic particles and 
developing the coherent-state path integral from the master equation 
\cite{cardyrv}, one can obtain the proper action $S$ which can be treated by 
the systematic many-body analysis like renormalization group 
(RG) calculation \cite{L94}.
Including biased diffusion (drift), the action for the pair annihilation/coagulation model is given as
\begin{equation}
{\cal S} = \int dt d {\bm x}\left[\bar \varphi ( \partial_t -D\nabla^2 + {\bm v}\cdot \nabla)
\varphi +  \lambda_1 \bar \varphi \varphi^2 + \lambda_2 \bar \varphi^2 \varphi^2\right],
\label{Eq:AAaction}
\end{equation}
where $D$ is the diffusion constant and ${\bm v}$ is the drift velocity, while
$\lambda_1$ and $\lambda_2$ are properly scaled reaction parameters. The particle density field
is denoted by $\varphi$ and its response field by $\bar \varphi $. 
The driving term can be simply gauged away by a Galilean transformation 
such as $\varphi(t,{\bm x}) \rightarrow \varphi(t, {\bm x} - {\bm v} t)$ and
$\bar \varphi(t,{\bm x}) \rightarrow \bar \varphi(t, {\bm x} - {\bm v} t)$.
Therefore, one can conclude that the driving is irrelevant for the pair annihilation/coagulation 
model in the long time regime. 

The argument based on the Galilean invariance can be applied to more general SRDA 
exhibiting absorbing phase transitions. Near the transition, the particle density is low enough to 
assure the validity of  the bosonic field theory. Only exceptions are found in some multi-species 
diffusion-reaction systems, where the hard core exclusion becomes crucial~ \cite{AB,KLP00}. 
The DP class is well known to be described by a single-species bosonic field 
theory as well as the PC class. Therefore, one can expect that the external driving does not
change the critical scaling. 

To confirm our expectation, we study the driven branching annihilating random walks 
with one (DBAW1) and two (DBAW2) offspring in one dimension. The models without
the external driving, the BAW1 and the BAW2,  belong to the DP and the PC class, respectively. 
The evolution dynamics for the driven models with fully biased diffusion
are summarized using stoichiometric notations as
\begin{equation}
\begin{aligned}
&A \emptyset \stackrel{p}{\longrightarrow} \emptyset A,
\quad
A A \stackrel{p}{\longrightarrow} \emptyset \emptyset,\\
&\begin{cases}
A\emptyset \stackrel{1-p}{\longrightarrow} AA, &\text{for one offspring},\\
A\emptyset\emptyset \stackrel{1-p}{\longrightarrow} AAA. &\text{for two offsprings}.
\end{cases} 
\end{aligned}
\end{equation}
For simplicity, the branching process is also taken to be biased, but
this choice does not change our conclusion.

\begin{figure}[b]
\includegraphics[width=0.4\textwidth]{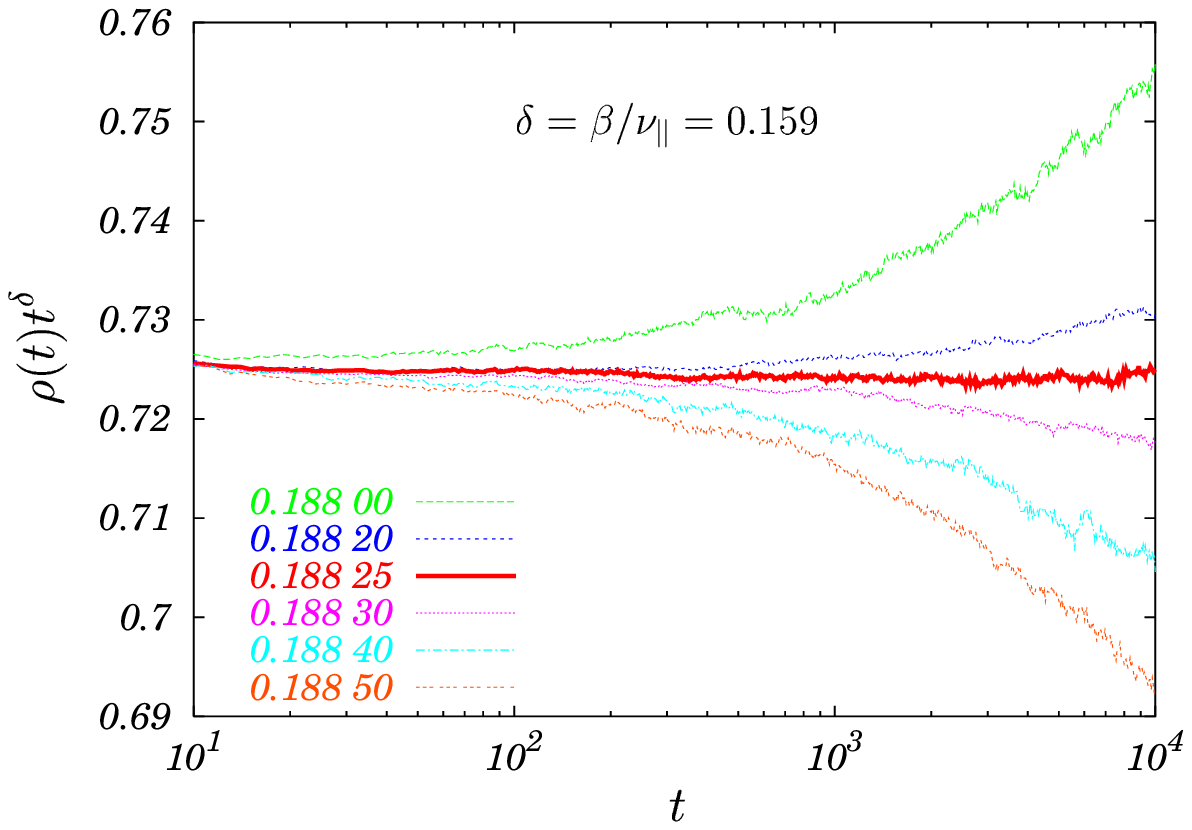}
\includegraphics[width=0.4\textwidth]{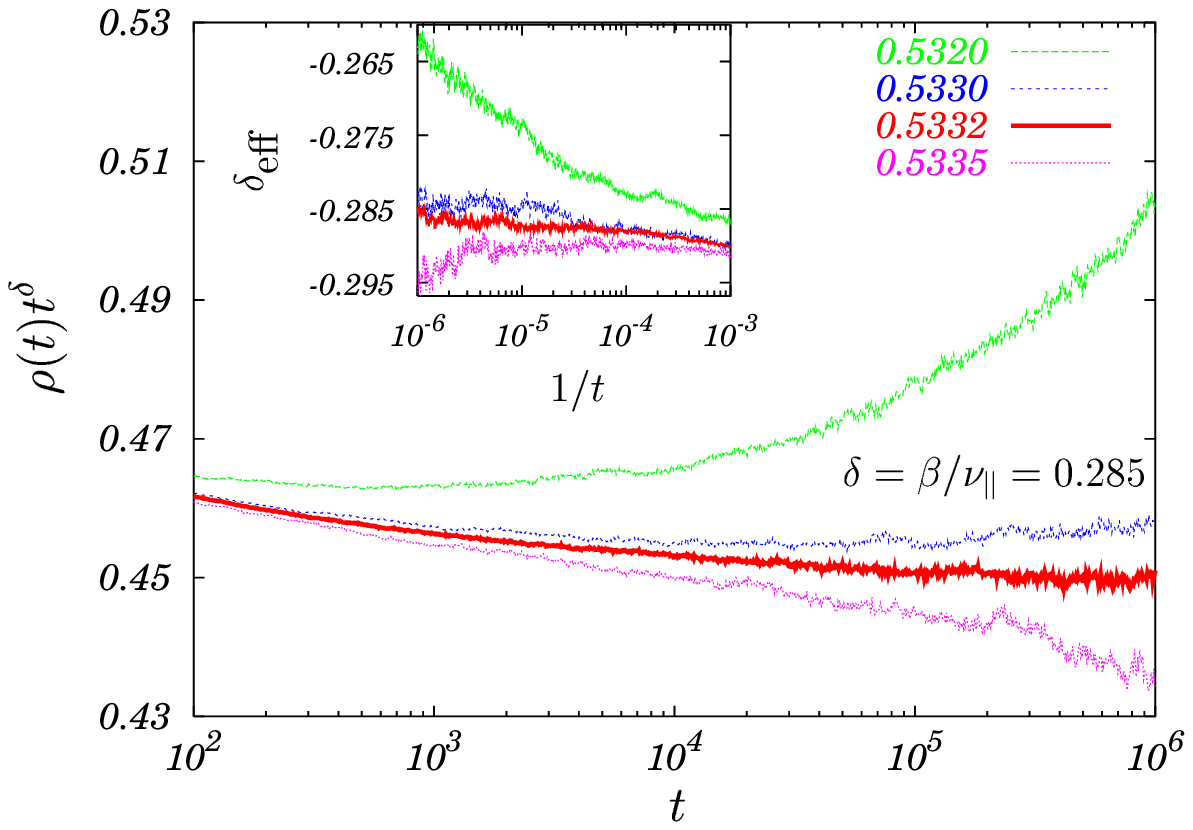}
\caption{\label{Fig:DP} (color online) Semi-log plots of $\rho(t) t^{\delta}$ 
vs $t$  for DBAW1 (upper panel) and DBAW2 (lower panel).
We use $\delta=0.159$ for DBAW1 and $0.285$ for DBAW2.
Since DBAW2 shows a long-term correction to scaling,  
we also draw the effective
exponents in the inset of the lower panel and 
find $\delta \simeq 0.285 (1)$ 
for DBAW2 which is consistent with the PC value.}
\end{figure}

We perform Monte Carlo simulations starting with the fully occupied initial condition.
The particle density $\rho(t)$ is measured as a function of time $t$ in a lattice of size   
$L = 2\times 10^5$ and $L = 10^6$ for the DBAW1 and the DBAW2,
respectively. Up to the observation time, all samples are alive
in our simulations.  Since a power law decay 
$\rho(t) \sim t^{-\delta}$ is expected at criticality, one should look
for a flat line in the $\rho(t) t^{\delta}$ vs $t$ plot to locate the
critical point. 
In Fig.~\ref{Fig:DP}, we find $p_c=0.18825(5)$ with $\delta = 0.159(1)$ for DBAW1 and
$p_c = 0.5332(2)$ with $\delta = 0.285(1)$ for DBAW2.
The values of the critical exponent ratios agree perfectly well with the known values for the DP and
the PC class. The driven systems with partial bias also show the same critical behaviors.
This is exactly what we expected from the Galilean invariance argument for the
SRDA. 

Now, we turn to the PCPD model and study the effect of driving on its critical scaling.
The model dynamics consists of three configurational changes such as (biased) diffusion
($A\emptyset\leftrightarrow\emptyset A$), pair annihilation ($2A\rightarrow \emptyset$), 
and creation of a particle by a pair ($2A\rightarrow 3A$).
The algorithm to simulate the driven PCPD (DPCPD) in one dimension is as follows:
First, choose a particle at random. The chosen particle attempts to hop to the right or to the left
with probability $D$ and $1-D$, respectively. If the target site is vacant, the hopping trial is accepted.
If the target site is occupied, (a) two particles annihilate with probability $p$ or 
(b) the hopping attempt is rejected and the pair (chosen particle and one at the target site) tries to
create a particle at a randomly chosen nearest neighbor site of the pair. When the selected site is
occupied, this branching attempt is rejected. The time increases by $1/N(t)$, where $N(t)$
is the total number of particles at time $t$.

\begin{figure}[b]
\includegraphics[width=0.4\textwidth]{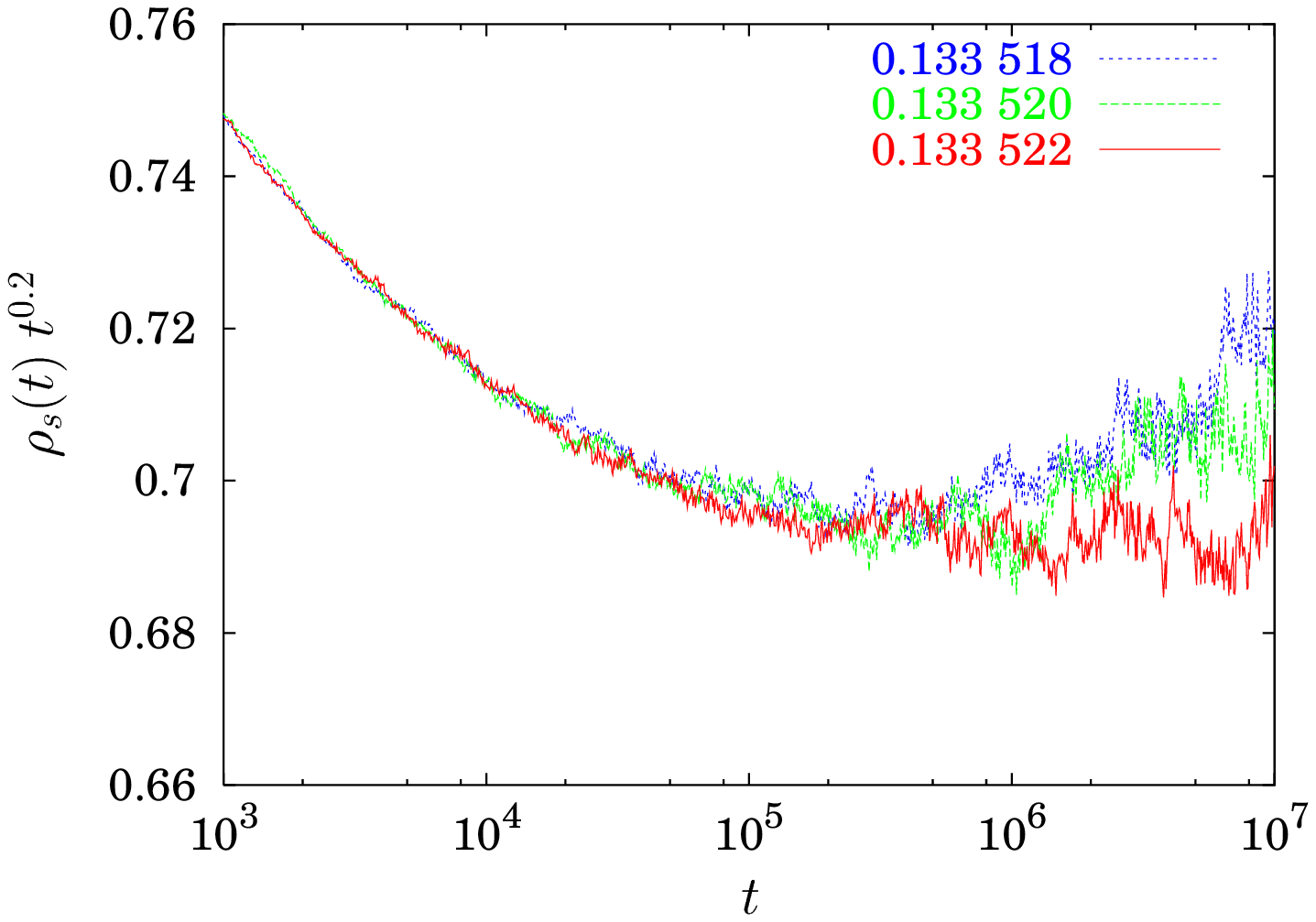}
\includegraphics[width=0.4\textwidth]{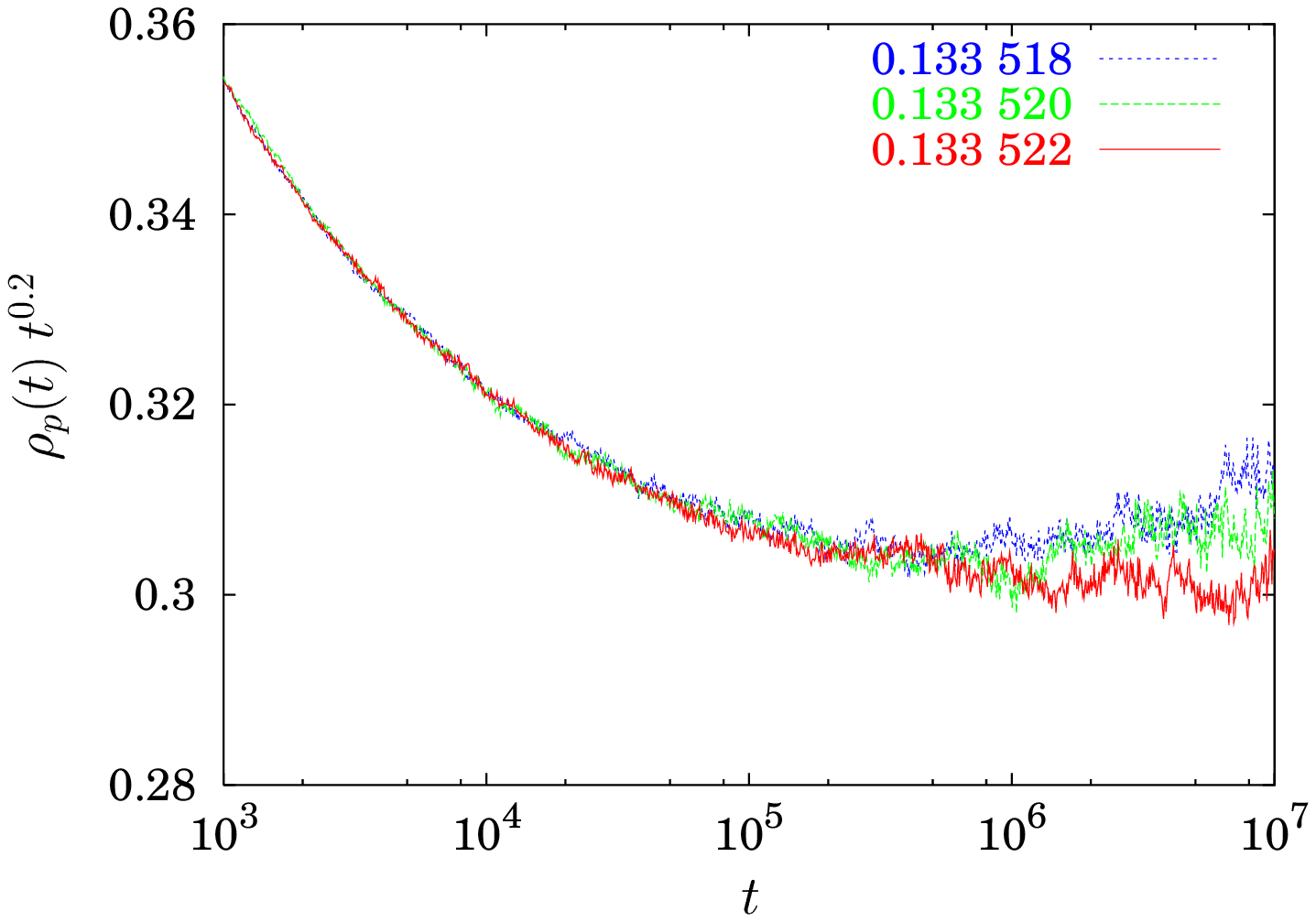}
\caption{\label{Fig:PCPD}(color online)  A semi-log plot of $\rho(t) t^{\delta}$
vs $t$  for the ordinary PCPD with normal diffusion ($D = 1/2$). 
In the upper (lower) panel, the data for the particle (pair) density are 
plotted.  We find a good flat line at criticality with $\delta = 0.20$.}
\end{figure}

We measure the particle density $\rho_s(t)$ and the nearest neighbor 
pair density $\rho_p(t)$ in a lattice of
size $L=10^7$ up to $t=10^8$ and average over $\sim 80$ independent samples.
At $D=1/2$, the ordinary PCPD with normal diffusion is recovered. 
In Fig. \ref{Fig:PCPD}, after a huge crossover time around $t\simeq 10^5$,
we see a flat straight line at criticality ($p_c = 0.133522(2)$) with 
$\delta = 0.20(1)$  for both particle and pair densities, which is 
in good agreement with the most reliable value for the PCPD~\cite{KC03}.

\begin{figure}[b]
\includegraphics[width=0.4\textwidth]{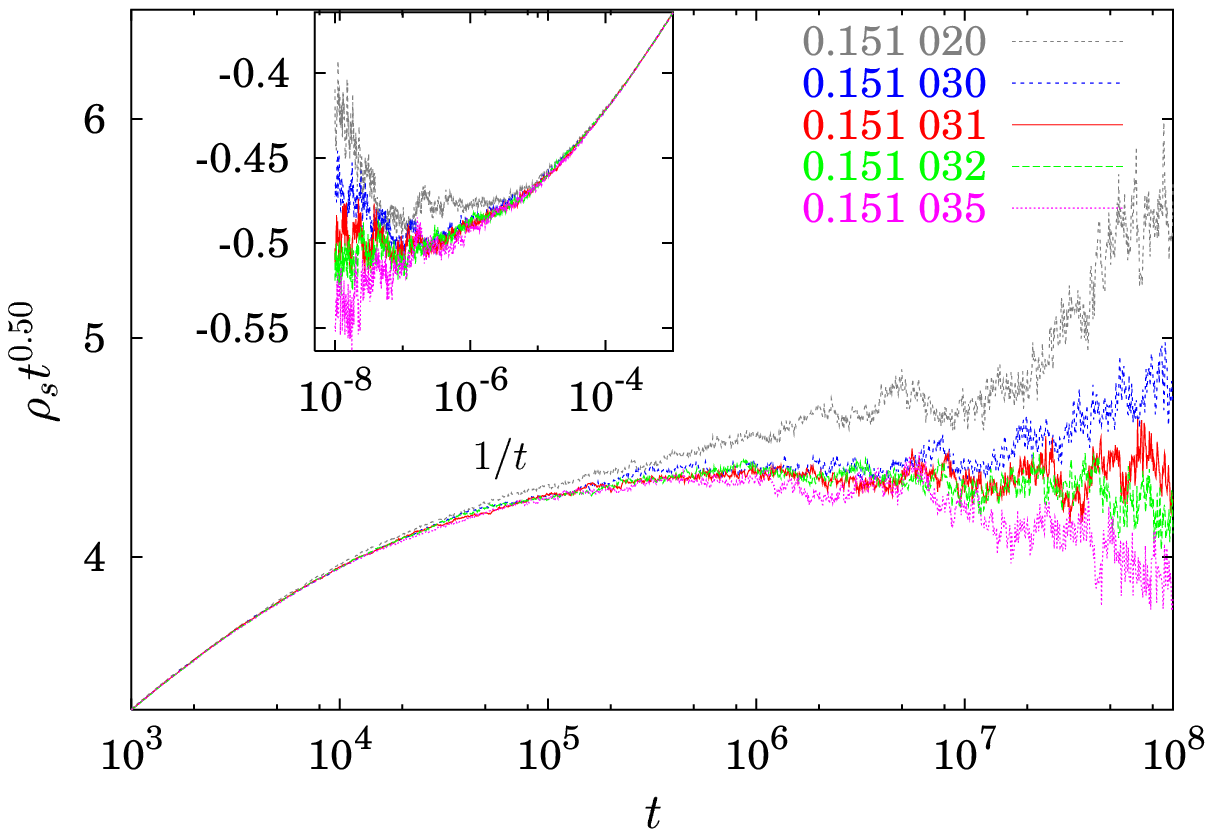}
\includegraphics[width=0.41\textwidth]{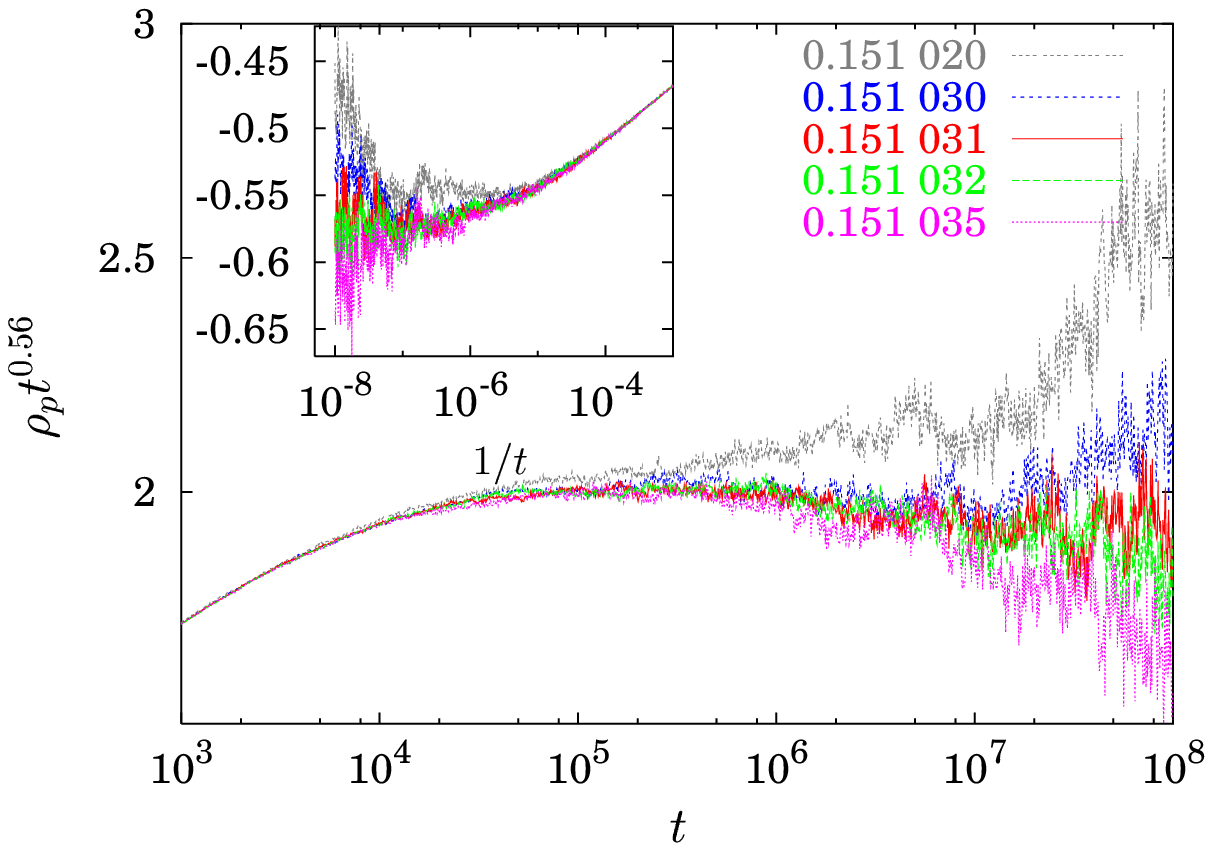}
\caption{\label{Fig:DPCPD_half}(color online)  Plots of $\rho_s(t) t^{0.49}$ 
vs $t$ for the particle density  and $\rho_p (t) t^{0.56}$ vs $t$ for the
pair density of the DPCPD model. In the inset of each panel, the effective
exponents are drawn as a function of $1/t$.}
\end{figure}

To see the effect of driving in the PCPD, we perform simulations at $D=1$ (full bias).
In Fig.~\ref{Fig:DPCPD_half}, we find $p_c = 0.151031(1)$ with 
$\delta_s = 0.49(1)$ for the particle density and $\delta_p = 0.56(3)$
for the pair density, which are unambiguously distinct from the value of the
ordinary PCPD, $\delta\simeq 0.20$. These results do not change for any partial bias.
This is a big surprise because it implies that the ordinary Galilean invariance 
should not hold in the PCPD under driving, which in turn can not  be described 
by a single-species bosonic field theory.

\begin{figure}[t]
\includegraphics[width=0.4\textwidth]{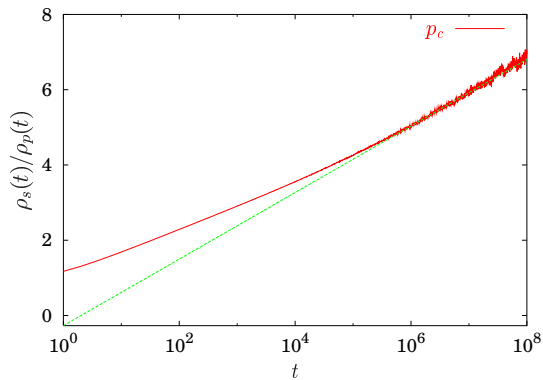}
\caption{\label{Fig:DPCPD_log}(color online)  The semi-log plot of 
$\rho_s/\rho_p$ vs $t$  at criticality. The straight line stands for the logarithmic
fitting of the data, which seems very good for nearly three decades.}
\end{figure}

Before going into detailed discussion on its implication, we note that 
there is another surprise that the exponent values are almost identical to 
the values of the ordinary PCPD in two dimensions (``mean-field" values)~\cite{OMS}. 
The upper critical dimension of the PCPD is expected to be two
and the decay dynamics presumably carries a multiplicative logarithmic 
factor in two dimensions.  We plot $\rho_s (t)/\rho_p (t)$ versus
$t$ in a semi-log scale in Fig.~\ref{Fig:DPCPD_log}, as in the 2D case
studied in~\cite{OMS}. It seems to confirm that
$\rho_s (t)/\rho_p (t)\sim \ln t$. Therefore, the critical scaling of $\rho_s$
and $\rho_p$ exhibits exactly the same critical behavior found at the 2D PCPD 
criticality. This may suggest that the upper critical dimension of the DPCPD
is 1 rather than 2. The reduction of the upper critical dimension by the 
biased diffusion is not rare. The most prominent example is the sandpile model 
related to the self-organized criticality. It is well known that the directed 
toppling rules lower the upper critical dimension from 4 to 3 \cite{DR89}. 
However, the situation is not so simple here.  The decay dynamics inside 
the absorbing phase remains to be one dimensional, i.e.~$\rho_s\sim t^{-1/2}$ 
and $\rho_p \sim t^{-3/2}$. Therefore, only the critical scaling carries the 2D
character, while the absorbing phase is of 1D characteristic.  The underlying 
mechanism for this surprising scaling behavior is under investigation.

Now, we come back to the implication given by the relevancy of the external 
driving.  It implies that the PCPD under driving can not be described by 
a single-species bosonic field theory. This reminds us of the interpretation 
of the PCPD as a cyclically coupled DP and annihilation process suggested by 
Hinrichsen~\cite{H01}, where a pair and a solitary particle can be considered 
two independent excitations (two-species particles). If we accept that these 
two excitations are independent, a field theory of two-species is more suitable
for the description of the PCPD. Then, the difference in the bias strength 
(drift velocity) for two different particles may be relevant as in the 
well-known two-species annihilation model $A+B\rightarrow\emptyset$~\cite{KR84}.
By introducing the biased diffusion of a single particle in the PCPD, the 
effective diffusion of a pair will be also biased but the drift velocity should 
be in general different each other. Therefore, our results suggest that the 
bias difference between two excitations are the reason for the relevancy of the
driving in the PCPD in the context of a two-species reaction diffusion model.

In order to understand this feature more clearly, we study the full bosonic
model with a {\em soft} constraint introduced by Kockelkoren and 
Chat\`e~\cite{KC03}, which belongs to the PCPD class. It is obvious that the 
biased diffusion does not change the critical scaling in this full bosonic 
model due to the Galilean invariance. However, this is very fortuitous. 
Once we apply the different diffusion bias to a particle at singly occupied 
sites and a particle at multiply occupied sites, we recover the mean-field 
exponents again~\cite{PPun}.  This confirms the role of 
the bias difference as a relevant perturbation to the PCPD fixed point.

We emphasize that the the bias difference is irrelevant for the multi-species 
models belonging to the DP class, because the DP is generically a 
single-species model.  To check it explicitly, we study the generalized PCPD 
(GPCPD) model introduced by Noh and Park~\cite{NP04}, which is parametrized
by the memory strength $r$. At $r=1$, the PCPD model is recovered, while the 
DP class is found at $r=0$. With biased diffusion, we find the mean-field 
exponents with logarithmic corrections for any finite $r$, but the DP is 
robust against this external driving at $r=0$~\cite{PPun}.  This again 
confirms that the PCPD (in general, GPCPD at nonzero $r$) should not belong 
to the DP class.

In conclusion, we studied the effect of bias on the critical scaling  in one-dimensional 
reaction-diffusion models. The BAW models are robust against the external driving, 
regardless of the parity conservation. This is  anticipated from the fact that
the DP and the PC class can be generically described by a single-species bosonic field theory,
where the Galilean invariance is embedded. In contrast, the driving is relevant for 
the PCPD and changes the critical scaling. This leads us to exclude a possibility of
the DP or the PC class for the critical scaling of the PCPD model. Moreover, it suggests that
the PCPD is generically a two-species model and a field theory of two-species may be required.

We thank J.~D.~Noh for useful discussions.

\end{document}